\documentclass[aps,pre,onecolumn,noshowpacs,superscriptaddress,groupedaddress,preprintnumbers,nofootinbib,showpacs,12pt]{revtex4}
\usepackage{epsfig,dsfont,amssymb,amsmath,amsthm,amsfonts,amsbsy,mathrsfs}
\usepackage{lipsum}
\usepackage{color}
\usepackage{amsmath}
\usepackage{amssymb}
\usepackage{subfigure}
\usepackage{float}
\usepackage{hyperref}
\usepackage{verbatim}
\usepackage{soul}
\setstcolor{red}
\newcommand {\be}{\begin{equation}}
\newcommand {\ee}{\end{equation}}
\newcommand {\bea}{\begin{eqnarray}}
\newcommand {\eea}{\end{eqnarray}}
\begin{document}

\title{How neural networks find generalizable solutions: Self-tuned annealing in deep learning}
\author{Yu Feng}
\affiliation{IBM T. J. Watson Research Center, Yorktown Heights, NY10598}
\affiliation{Department of Physics, Duke University, Durham, NC27710}
\author{Yuhai Tu}
\affiliation{IBM T. J. Watson Research Center, Yorktown Heights, NY10598}
\email{yuhai@us.ibm.com}

\begin{abstract}

Despite the tremendous success of Stochastic Gradient Descent (SGD) algorithm in deep learning, little is known about how SGD finds generalizable solutions in the high-dimensional weight space. By analyzing the learning dynamics and loss function landscape, we discover a robust inverse relation between the weight variance and the landscape flatness (inverse of curvature) for all SGD-based learning algorithms. To explain the inverse variance-flatness relation,  we develop a random landscape theory, which shows that the SGD noise strength (effective temperature) depends inversely on the landscape flatness. Our study indicates that SGD attains a self-tuned landscape-dependent annealing strategy to find generalizable solutions at the flat minima of the landscape. Finally, we demonstrate how these new theoretical insights lead to more efficient algorithms, e.g., for avoiding catastrophic forgetting.

\end{abstract}      
\maketitle
\clearpage


\section{Introduction}

One key ingredient for the powerful deep neural network (DNN) based machine learning paradigm --Deep Learning~\cite{LeCun2015Deep}-- is a relatively simple iterative method called stochastic gradient descent (SGD)~\cite{Robbins_1951,SGD2010Bottou}. However, despite the tremendous successes of Deep Learning, the reason why SGD is so effective in learning in a high dimensional nonconvex loss function (energy) landscape remains poorly understood. The random element seems key for SGD, yet makes it harder to understand.  Fortunately, many physical systems include such random element, e.g., Brownian motion, and powerful tools have been developed for understanding collective behaviors in stochastic systems with many degrees of freedom. Here, we use concepts and methods from statistical physics to investigate the SGD dynamics, the loss function landscape, and more importantly their relationship. 

We start by introducing the SGD based learning process as a stochastic dynamical system. A learning system such as neural network (NN) especially DNN has a large number ($N$) of weight parameters $w_i$ $(i=1,2,...,N)$. For supervised learning, there is a set of $M$ training samples each with an input $\vec{X}_k$ and a correct output $\vec{Z}_k$ for $k=1,2,...,M$. For each input $\vec{X}_k$, the learning system predicts an output $\vec{Y}_k=G(\vec{X}_k,\vec{w})$, where the output function $G$ depends on the architecture of the NN as well as its weights $\vec{w}$. The goal of learning is to find the weight parameters to minimize the difference between the predicted and correct output characterized by an overall loss function (or energy function):
\begin{equation}
    L(\vec{w})=M^{-1}\sum_{k=1}^{M} d(\vec{Y}_k,\vec{Z}_k),
\end{equation}
where $d(\vec{Y}_k,\vec{Z}_k)$ is a measure of distance between $\vec{Y}_k$ and $\vec{Z}_k$. In our study, a cross-entropy loss for $d$ (see Supplementary Material (SM) for its expression) is used . 

One learning strategy is to update the weights by following the gradient of $L$ directly. However, this batch learning scheme is computationally inhibitive for large datasets and it also has the obvious shortfall of being trapped by local minima. SGD was first introduced to circumvent the large dataset problem by updating the weights according to a subset (minibatch) of samples randomly chosen at each iteration ~\cite{Robbins_1951}. Specifically, the change of weight $w_i$ $(i=1,2,...,N)$ for iteration $t$ in SGD is given by:
\begin{equation}
\Delta w_i (t) = -\alpha \frac{\partial L^{\mu(t)}(\vec{w})}{\partial w_i},
\label{backprop}
\end{equation}
where $\alpha$ is the learning rate and $\mu(t)$ represents the random minibatch used for iteration $t$. The mini loss function (MLF) for minibatch $\mu$ of size $B$ is defined as:
\begin{equation}
    L^\mu(\vec{w}) =B^{-1} \sum_{l=1}^{B} d(\vec{Y}_{\mu_l},\vec{Z}_{\mu_l}),
\end{equation} 
where $\mu_l$ ($l=1,2,..,B$) labels the $B$ randomly chosen samples. 

Here, we introduce the key concept of a MLF ensemble $\{L^\mu(\vec{w})\}$, i.e., an ensemble of energy landscapes each from a random minibatch. The overall loss function $L(\vec{w})$ is just the ensemble average of MLF:
$L \equiv \langle L^\mu\rangle _\mu$. The SGD noise comes from the variation between a MLF and its ensemble average:
$\delta L^{\mu} \equiv L^{\mu}-L$.
By taking the continuum time limit in Eq.~\ref{backprop}, we obtain the following stochastic partial differential equation for SGD:
\begin{equation}
\frac{\partial \vec{w}}{\partial t}
= -\alpha \frac{\partial L}{\partial \vec{w}}+\vec{\eta}(\vec{w}),
\label{Langevin}
\end{equation}
where we have taken the learning step as the unit of time. This equation is analogous to the Langevin equation in statistical physics. The first term $-\alpha \frac{\partial L}{\partial \vec{w}}$ is the deterministic gradient descent governed by the overall loss function $L$ analogous to the energy function in physics. The second term is the SGD noise term 
$
\vec{\eta}\equiv  -\alpha \nabla_{\vec{w}} \delta L^{\mu}(\vec{w})
$
with zero mean 
$\langle \vec{\eta} \rangle = 0$
and equal time correlation 
$C_{ij}(\vec{w})\equiv \langle \eta_{i} \eta_{j} \rangle = \alpha^2  \times \langle \frac{\partial \delta L^{\mu}}{\partial w_i} \; \frac{\partial \delta L^{\mu}}{\partial w_j} \rangle _\mu$, which depends explicitly on $\vec{w}$. 

As first pointed out by Chaudhauri and Soatto~\cite{Chaudhari_2018}, unlike equilibrium physical systems where the noise has a constant strength given by the thermal temperature, the SGD dynamics is highly nonequilibrium as the SGD noise is anisotropic and varies in the weight space. 
In the rest of the paper, we study the SGD-based machine learning process by adopting the nonequilibrium stochastic dynamics framework. Our working hypothesis is that SGD may serve as an efficient annealing strategy for varying the noise (or effective temperature) ``intelligently" according to the loss function landscape in order to find the shallow (flat) minima where ``good" (generalizable) solutions are believed to exist~\cite{Hinton1993Keeping,hochreiter1997flat,BaldassiE7655,chaudhari2016entropysgd,Zhang_2018,Mei2018MeanField}.

\section{Learning via a low-dimensional drift-diffusion dynamics in SGD}

In general, SGD based DNN learning dynamics can be roughly divided into an initial fast learning phase where $L$ decreases rapidly, followed by an ``exploration" phase where the weights keep evolving but the loss $L$ only changes very slowly (see Fig. S1 in SM). The weight vectors sampled in the exploration phase are treated as solutions to the problem given their low losses.

We used principal component analysis (PCA) to study the stochastic motion of the weight vector during the SGD learning process, in particular during the exploration phase when the system reaches a quasi-steady-state (see Methods section for details of PCA).  Within a large time window $t\in [t_0,t_0+T]$ where $t_0$ is a time in the exploration phase and $T(=10\; $epochs\footnote{each epoch has $\frac{M}{B}$ iterations which covers all training samples once} used here) is a large time window, the weight dynamics can be decomposed into its variations in different principal components:
\begin{equation}
\vec{w}(t)=\langle \vec{w}\rangle _T+\sum_{i=1}^{N} \theta_i(t)\vec{p}_i,
\end{equation}
where $\langle \vec{w}\rangle _T=T^{-1}\int_{t_0}^{t_0+T}\vec{w}(t)dt$ is the average weight vector in the time window $T$, $\vec{p}_i$ is the $i$-th principal component base vector with $\vec{p}_i\cdot \vec{p}_j=\delta _{ij}$. The projection of the weight vector along the PCA direction $\vec{p}_i$ is given by $\theta_i(t)$, which is a linear combination of the individual weights, $\vec{\theta}$ is the weight vector in the PCA coordinate. 

\begin{figure}[htbp]
\centering
\includegraphics[width=0.9\linewidth]{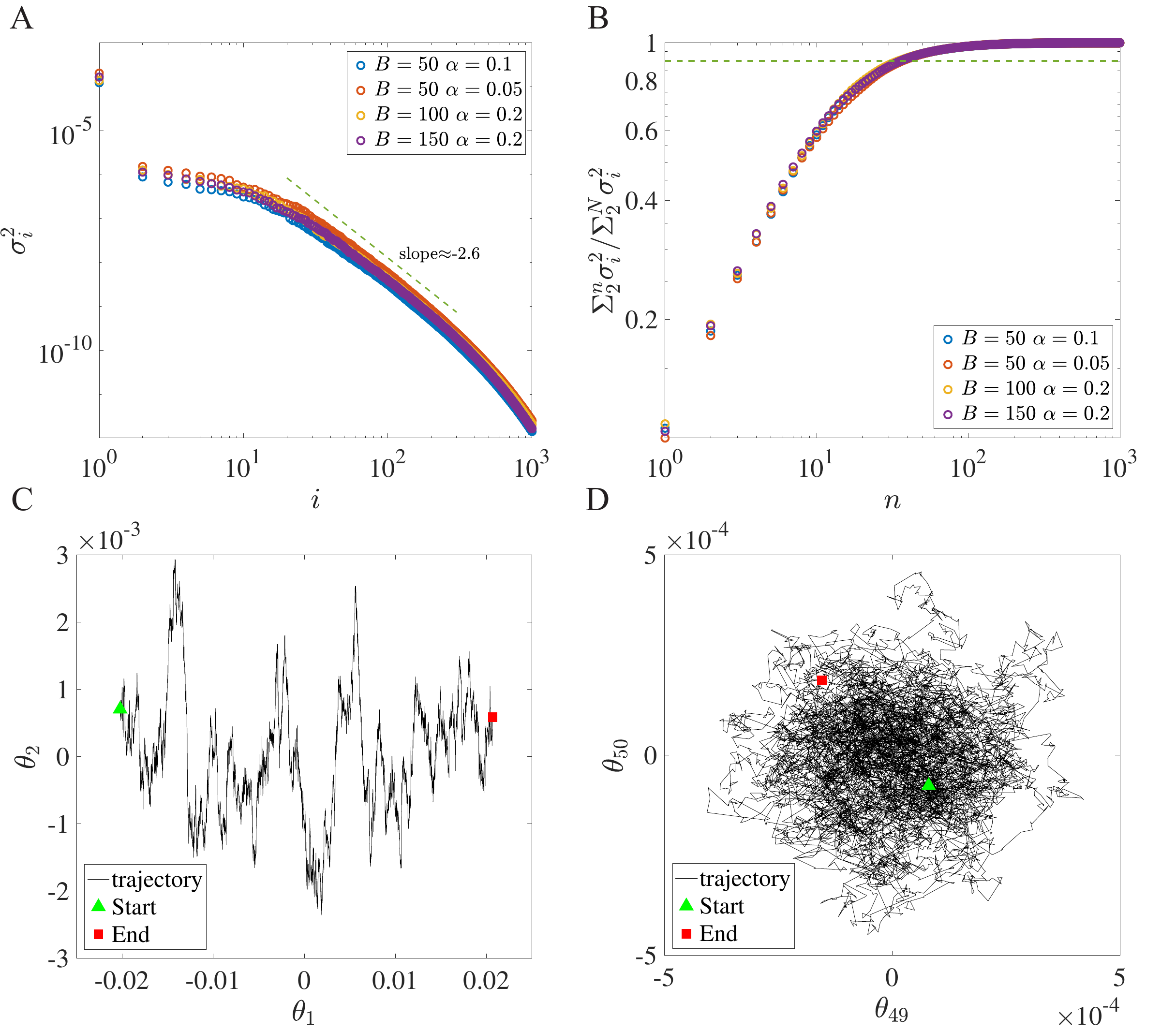}
\caption{The PCA results and the drift-diffusion motion in SGD. (A) The rank-ordered variance $\sigma_i^2$ in different principal component (PC) direction $i$. For $i\ge 20$, $\sigma_i^2$ decreases with $i$ as a power law $i^{-\gamma}$ with $\gamma \sim 2-3$. (B) The normalized accumulative variance of the top $(n-1)$ PCA components excluding $i=1$. It reaches $\sim 90\%$ at $n=35$ much smaller than the total number of weights $N=2500$ between the two hidden layers. (C) The SGD weight trajectory projected onto the $(\theta_1,\theta_2)$ plane. The persistent drift motion in $\theta_1$ and the diffusive random motion in $\theta_2$ are clearly shown. (D) The diffusive motion in $(\theta_i,\theta_{j})$ plane with $j>i(\ne1)$ randomly chosen ($i=49$ and $j=50$ shown here). Unless otherwise stated, hyperparameters used are: $B=50$, $\alpha=0.1$.}  
\label{fig:1} 
\end{figure}

The results reported here are for a simple NN with 2 hidden layers each with $50$ neurons for classification tasks using the MNIST database (see Methods for details and other NN architectures used). The PCA was done for the $N=2500$ weights between the two hidden layers (results for other NN architectures and databases are included in the SM). In Fig.~\ref{fig:1}A, we show the PCA spectrum, i.e., the variance $\sigma^2 _i\equiv T^{-1}\int_{t_0}^{t_0+T}\theta_i^2(t)dt$  versus its rank $i$ in descending order $(\sigma_{i+1}<\sigma_i)$.  We found that the variance in the first PCA direction ($\vec{p}_1$) is much larger than variances in other directions because the motion along $\vec{p}_1$ has a net drift velocity (see discussion below and Fig.~\ref{fig:1}C). For other PCA directions, after a small number of leading PCA directions $2\le i \le 20)$, the variance $\sigma^2_i$ decays rapidly with its rank:  $\sigma_i^2 \sim i^{-\gamma}$ for $20<i<200$ with a large exponent $\gamma\sim 2.6$ before an even faster decay for higher $i(>200)$. This means that most of the variations (dynamics) of the weights is concentrated in a relatively small number of PCA directions (dimensions). Quantitatively, as shown in Fig.~\ref{fig:1}B, even excluding $\sigma_1^2$, more than $90\%$ of the total variance occurs in the first $35$ PCA modes much smaller than the total number of weights $N=2,500$, which suggests that the SGD dynamics are embedded in a low dimensional space\cite{GurAir2018Gradient}. 

Next, we studied the network dynamics along different PCA directions. We found that along the first PCA direction $\vec{p}_1$, there is a net drift velocity $d\theta_1/dt$ with a persistence time much longer than $1$ epoch  as clearly shown in Fig.~\ref{fig:1}C where SGD dynamics projected onto the $(\theta_1,\theta_2)$ space is shown. For all other PCA directions, the dynamics are random walks with a short velocity correlation time (shorter than $1$ epoch) as clearly demonstrated in Fig.~\ref{fig:2}C where the SGD dynamics projected onto a randomly chosen pair of PCA directions $(\theta_{49},\theta_{50})$ is shown. 

The persistent drift in the first PCA direction $\vec{p}_1$ can be understood by moving a solution $\vec{w}_0$ found by SGD  along $\vec{p}_1$ by $\theta_1$ to a new weight vector $\vec{w}=\vec{w}_0+\theta_1 \vec{p}_1$. We find that such a translation results roughly in an overall amplification of 
the difference between the outputs for the right class and the wrong classes, which leads to a change in the cross-entropy loss function $L(\vec{w})\approx L(\vec{w}_0)\times \exp(\beta \theta_1)$ with $\beta$ a constant parameter.  This dependence of $L$ on $\theta_1$ leads to the persistent motion along the $\vec{p}_1$ direction with a low speed proportional to $L$ which slowly decreases with time itself (see SM for details). In the next section, we focus on the majority diffusive PCA modes ($i\ge 2$) and their relation to the loss function landscape. 

\section{The loss function landscape and the inverse variance-flatness relation}

In the exploration phase, the loss function is small and all the weight vectors along the SGD trajectory can be considered as valid solutions.  However, the solutions found by a SGD trajectory only represent a small subset of valid solutions. To gain insights on the full solution space, we study the loss function landscape around a specific solution $\vec{w}_0$ reached by SGD. Specifically, we compute the loss function profile $L_i$ along the $i$-th PCA direction $\vec{p}_i$ determined from PCA of the SGD dynamics:
\begin{equation}
L_i(\delta\theta) \equiv L(\vec{w}_0+\delta \theta \vec{p}_i).
\end{equation} 

\begin{figure}[htbp]
\centering
\includegraphics[width=0.9\linewidth]{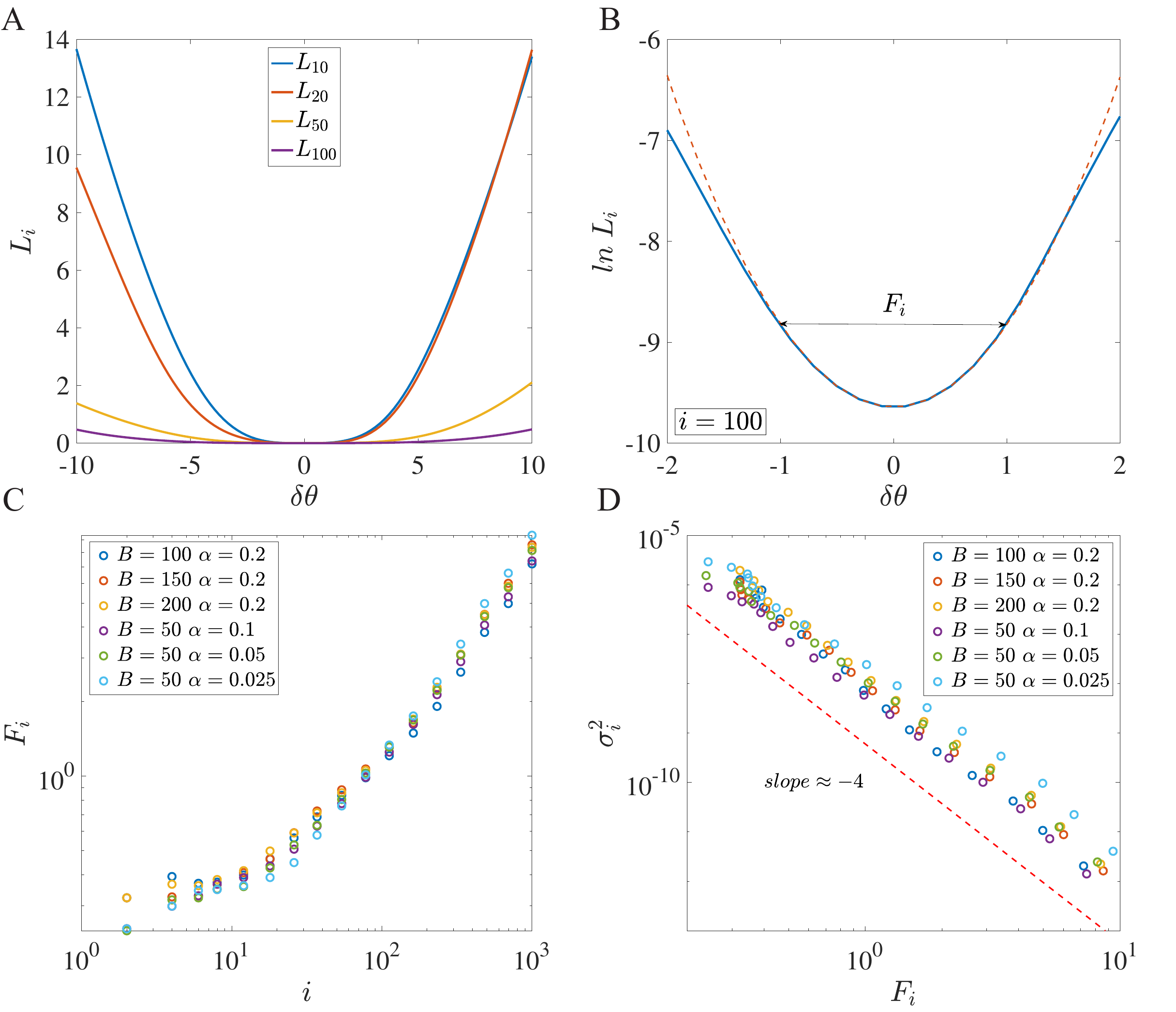}
\caption{The loss function landscape and the inverse variance-flatness relation. (A) The loss function profile $L_i$ along the $i$-th PCA direction. (B) The $\ln(L_i)$ profile. It can be fitted by a quadratic function (the red dotted line). The definition of the flatness $F_i$ is also shown. (C) The flatness $F_i$ for different PCA direction $i$, and (D) the inverse relation between the variance $\sigma^2_i$ and the flatness $F_i$ for different choices of minibatch size $B$ and learning rate $\alpha$.} 
\label{fig:2} 
\end{figure}

In Fig.~\ref{fig:2}A, we show the loss function landscape profiles $L_i(\delta\theta)$ for several diffusive PCA directions $i=10,20,50,100$. Surprisingly, the loss function profiles are smooth and flat in all directions.
The flat landscape can be seen more clearly in Fig.~\ref{fig:2}B where we plot $\ln(L_i(\delta\theta))$ versus $\delta\theta$, which shows that $\ln(L_i(\delta\theta))$ can be fitted well by a quadratic function of $\delta\theta$ near $\delta\theta=0$ for all $i\ge 2$ (see Fig.~S2 in SM for the landscape for $i=1$): 
\begin{equation}
\ln (L_i(\delta\theta))\approx \ln (L_0) + \frac{4\delta\theta^2}{F_i^2},
\label{land}
\end{equation}
where $L_0\equiv L(\vec{w}_0)$ is the loss at $\vec{w}_0$ and $F_i$, inversely proportional to the curvature, defines a flatness parameter of the loss function landscape in the $i$-th PCA direction. In general, $F_i$ can be defined as the width of the region within which $L_i\le e\times L_0$. A larger value of $F_i$ means a flatter landscape in the $i$-th PCA direction.

Quantitatively, we found that the flatness $F_i$ increases with $i$ as shown in Fig.~\ref{fig:2}C. Given that the SGD variance $\sigma_i^2$ decreases with $i$ as shown in Fig.~\ref{fig:2}A, this immediately suggests an inverse relationship between the loss function landscape flatness and the SGD variance. Indeed, as shown in Fig.~\ref{fig:2}D, the inverse variance flatness relation holds true for different choices of minibatch size $B$ and learning rate $\alpha$. Quantitatively, the variance-flatness follows approximately a power law: 
\begin{equation}
     \sigma_i^2 \sim F_i ^{-\psi},
 \label{SF}
\end{equation}
where the exponent $\psi \sim 4 $ for different choices of $B$ and $\alpha$. 

The inverse variance-flatness relation is counter-intuitive. In equilibrium systems, the fluctuation of a variable around its equilibrium value is proportional to the change of the variable in response to an external perturbation, which is known as the fluctuation-response (or fluctuation-dissipation) relation~\cite{forster2018hydrodynamic}. Since the response is proportional to the flatness of the energy function or equivalently the loss function used in machine learning, the fluctuation-response relation in equilibrium systems means that the variance of a variable should be larger for a flatter landscape, which is the opposite to the inverse relation shown in Fig.~\ref{fig:2}D for SGD. 

What is the reason for the inverse variance-flatness relation in SGD? Unlike equilibrium systems where the noise strength (temperature) is a constant, the SGD noise comes from the difference between gradient of a random MLF and that of the overall (mean) loss function and thus it varies in the weight space and in time. 
In the next section, we explain the inverse variance-flatness relation based on the dependence of the SGD noise on the statistical properties of the MLF ensemble. 

\section{The random landscape theory and origin of the inverse variance-flatness relation}

The most distinctive feature of SGD is that at any given iteration (time) the learning dynamics is driven by a random minibatch out of an ensemble of minibatches each with its own random mini loss function (MLF). To understand the SGD dynamics, we develop a random landscape theory to describe the statistical properties of the MLF ensemble near a solution $\vec{w}_0$ (we set $\vec{w}_0=0$ for convenience). 

\begin{figure}[htbp]
\centering
\includegraphics[width=0.9\linewidth]{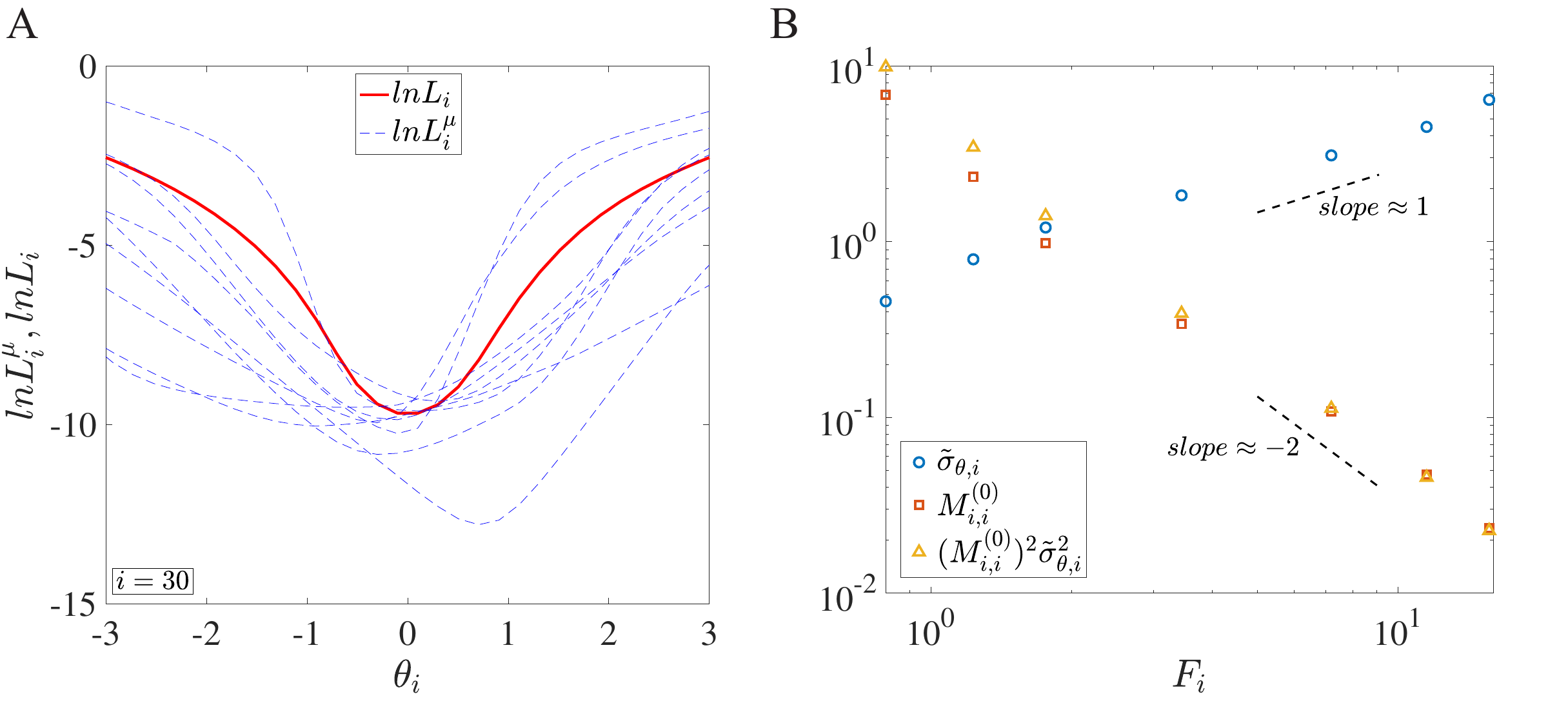}
\caption{Statistical properties of the MLF ensemble. (A) Profiles of the overall loss function $\ln(L_i)$ (red line) and a set of randomly chosen MLFs $\ln(L^\mu_i)$ (blue dashed lines) in a given PCA direction $i$. (B) The variance $\tilde{\sigma}^2_i$ of the minimum positions and the average diagonal element $M^{(0)}_{ii}$ of the Hessian matrices of the MLF ensemble versus the flatness $F_i$ of the overall loss function. The combination $(M^{(0)}_{ii})^2 \tilde{\sigma}^2_i$ versus $F_i$ is also shown. $i=30$ used here.}  
\label{fig:3} 
\end{figure}

As shown in Fig.~\ref{fig:3}A, we found that $L^\mu$ has roughly the same general shape as $L$ near its own minimum. Therefore, we approximate $L^\mu$ by an inverse Gaussian function with random parameters:
\begin{eqnarray}
L^\mu (\vec{\theta})  &\approx & L_{min}^\mu \exp[\sum_{i,j=1}^{N} \frac{M^\mu_{ij}}{2} (\theta_i -\theta_i^{\mu})(\theta_j -\theta_j^{\mu})]\nonumber \\ &=& L_{0}^\mu \exp[\sum_i \frac{M^\mu_{ii}}{2} \theta_i(\theta_i-2\theta_i^\mu) +\sum_{i<j} M^\mu_{ij} (\theta_i \theta_j -\theta_i \theta_j^{\mu}-\theta_j \theta_i^\mu)] \label{Lmu},
\end{eqnarray}
where $\vec{\theta}^\mu$ is the location of the minimum for MLF $L^\mu$, $\theta_i =\vec{\theta}\cdot \vec{p}_i$ is the parameter vector projected onto the $i$-th PCA direction, $L_{min}^\mu$ is the minimal loss, and ${\bf M}^\mu=\{M^\mu _{ij}\}$ is the symmetric Hessian matrix for $\ln(L^\mu)$ at its minimum location $\vec{\theta}^\mu$. 
For convenience, we define $L_0^\mu \equiv L^\mu (0)=L_{min}^\mu \exp[\sum_{i,j=1}^N M^\mu_{ij}\theta_i^\mu \theta_j^\mu]$ to represent the loss function value for minibatch $\mu$ at $\vec{\theta}=0$, and $L_0\equiv \langle L_0^\mu \rangle_\mu$ is the minimum loss of the overall loss function $L$.  
 
The MLF ensemble is described by the joint distribution of the parameters $M^\mu_{ij}$, $M^\mu_{ii}$, $\theta_i^\mu$, and $L_0^\mu$. By making the mean-field approximation that these parameters are uncorrelated random variables with Normal distributions: $M_{ii}^\mu\sim \mathcal{N}(M_{ii}^{(0)},\sigma_{M,i}^2)$ with positive mean $M_{ii}^{(0)}>0$, $M_{ij}^\mu\sim \mathcal{N}(0,\sigma_{ij}^2)$ with zero mean for $i\ne j$, $\theta_i^\mu\sim \mathcal{N}(0,\sigma_{\theta,i}^2)$ with zero mean for the diffusive modes ($i\ge 2$), and $\theta_1^\mu\sim \mathcal{N}(\theta_1^{(0)},\sigma_{\theta,1}^2)$ with a finite mean $\theta_1^{(0)}(\ne 0)$ for the drift mode ($i=1$), we obtain the overall loss function analytically (up to the second order terms in \textbf{$\theta_{i}$}) by averaging over the distributions of ${\bf M}^\mu$ and $\vec{\theta}^\mu$:
\begin{eqnarray}
L \equiv \langle L^\mu(\vec{\theta})\rangle_\mu &\approx& L_0 \big\langle \exp\{\sum_i \frac{M^\mu_{ii}}{2} \theta_i(\theta_i-2\theta_i^\mu) +\sum_{i<j} M^\mu_{ij} (\theta_i \theta_j -\theta_i \theta_j^{\mu}-\theta_j \theta_i^\mu)\}\big\rangle_{{\bf M}^\mu,\vec{\theta}^\mu}\nonumber \\
  &\approx & L_0\exp(-M_{11}^{(0)}\theta_1^{(0)}\theta_1+\sum_i \frac{M_{ii}}{2} \theta_i^2),
  \label{LMii}  
\end{eqnarray}
which has an inverse Gaussian form that is consistent with the empirical results shown in Fig.~\ref{fig:2}A\&B. The flatness $F_i \equiv (8/M_{ii})^{1/2}$ with $M_{ii}=M_{ii}^{(0)}+\sigma_{\theta,i}^2 (M_{ii}^{(0)})^2 + \sigma_{\theta,i}^2 \sigma_{M,i}^2 +\sum_{j\ne i} \sigma_{ij}^2 [ \sigma_{\theta,j}^2+\delta_{j1}(\theta_1^{(0)})^2]$, which depends on statistical properties of the MLF ensemble.

For each MLF $L^\mu$, if we vary $\vec{w}=\vec{w}_0+\theta_i \vec{p}_i$ along a given PC-direction $\vec{p}_i$, we obtain the MLF profile along the $i$-th PCA direction consistent with those shown in Fig.~\ref{fig:3}A from direct simulations:
$L^\mu_i (\theta_i) \propto \exp [M_{ii}^\mu (\theta_i-\tilde {\theta}_i^\mu)^2/2] $,
which has a minimum at $\theta_i= \tilde {\theta}_i^\mu$ that is shifted from $\theta_i^\mu$ due to the random off-diagonal Hessian matrix elements:
$   \tilde {\theta}_i^\mu = \theta_i^\mu +\frac{1}{M_{ii}^\mu}\sum_{j\ne i} M_{ij}^\mu$.  
It is easy to show that $\tilde {\theta}_i^\mu$ has zero mean ($\langle \tilde {\theta}_i^\mu\rangle_\mu=0$) and a variance given by:
\begin{equation}
    \tilde{\sigma}^2_{\theta,i} \equiv \langle (\tilde{\theta}_i^\mu)^2 \rangle_\mu = \sigma^2_{\theta,i} + \langle (M_{ii}^\mu)^{-2} \rangle_\mu \sum_{j\ne i}\sigma_{ij}^2 \sigma^2_{\theta,j}\approx \sigma^2_{\theta,i} + \frac{1}{(M_{ii}^{(0)})^2}\sum_{j\ne i}\sigma_{ij}^2 \sigma^2_{\theta,j},
    \label{tilde_sigma}
\end{equation}
where we have used the approximation $\langle (M_{ii}^\mu)^{-2} \rangle_\mu \approx \frac{1}{(M_{ii}^{(0)})^2}$. 

Given the explicit expression for $L^\mu$, Eq.~\ref{Lmu}, we can now study the SGD dynamics analytically. 
By keeping only up to the linear order in $\theta_i$, the SGD Langevin equation for $\theta_i$ becomes:
\begin{equation}
\frac{d\theta_i}{dt}=-\alpha \frac{\partial L^\mu}{\partial \theta_i}\approx  -\alpha L_0^\mu \sum_j M^\mu_{ij} (\theta_j-\theta_j^\mu),
\label{dthetadt1}
\end{equation}
which has an intuitive interpretation. At time $t$, the weight vector $\vec{\theta}$ is pulled by a random minibatch $\mu(t)$, whose MLF acts as a spring with a spring tensor ${\bf M}^\mu$ and its force center positioned at $\vec{\theta}^\mu$. The average relaxation time, $\tau_i \equiv 1/(\alpha L_0  M_{ii})$, is long given that $L_0$ is small. For $t\ll \tau_i$, variation of $\theta_j$ is much smaller than that of the minimum position $\theta^\mu_j$, i.e.,  $|\theta_j|\ll \sigma_{\theta,j}$. Therefore, we can neglect $\theta_j(\ll \theta_j^\mu)$ in Eq.~\ref{dthetadt1}, and we have:
\begin{equation}
\frac{d\theta_i}{dt}\equiv v_i =\alpha L_0^\mu \sum_j M^\mu_{ij} \theta_j^\mu,
\label{DD}
\end{equation}
which explains the drift-diffusion motion observed in our numerical simulations. For the first PCA direction ($i=1$), since $\langle \theta_1^\mu\rangle_\mu =\theta_1^{(0)}\ne 0$, there is a net drift velocity $v_1^{(0)}\equiv \langle v_1 \rangle_\mu \approx \alpha L_0 M_{11}^{(0)} \theta_{1}^{(0)}$. For all the other PCA directions ($i\ge 2$),  Eq.~\ref{DD} describes a diffusive motion with zero mean velocity $\langle v_i\rangle_\mu =0$ and a diffusion constant $D_i$ given by:
\begin{equation}
D_i \equiv \langle v_i^2\rangle \Delta t \approx  \alpha^2 \langle (L_{0}^\mu)^2\rangle_\mu [ (M_{ii}^{(0)})^2\sigma_{\theta,i}^2 +\sum_{j\ne i} \sigma^2_{ij}\sigma_{\theta,j}^2 ]  =\alpha^2 \langle (L_{0}^\mu)^2\rangle_\mu (M_{ii}^{(0)})^2\tilde{\sigma}_{\theta,i}^2,
\label{diff1}
\end{equation}
where we have used the expression for $\tilde{\sigma}_{\theta,i}$ from Eq.~\ref{tilde_sigma}.

The above result, Eq.~\ref{diff1}, provides a clear explanation for the inverse variance-flatness relation for the diffusive modes ($i\ge 2$). Qualitatively, a flatter landscape has a smaller value of $M_{ii}^{(0)}$, which leads to a smaller diffusion constant $D_i$ and thus a smaller variance since $\sigma_i^2 \propto D_i \propto (M_{ii}^{(0)})^2\tilde{\sigma}_{\theta,i}^2$. Quantitatively, both $M_{ii}^{(0)}$ and $\tilde{\sigma}^2_{\theta,i}$ can be determined from the MLF ensemble statistics. As shown in Fig.~\ref{fig:3}B, $\tilde{\sigma}_{\theta,i}$ scales almost linearly with $F_i=(8/M_{ii})^{1/2}$ and $ M^{(0)}_{ii}$ scales inversely with $F_i$, approximately as $F_i^{-2}$.
As a result, we have   
$\sigma_i^2 \propto  (M_{ii}^{(0)})^2\tilde{\sigma}_{\theta,i}^2 \propto F_i^{-2}$ (see Fig.~\ref{fig:3}B) which is consistent with the inverse power law dependence shown in Fig.~\ref{fig:2}D. However, the theoretically obtained exponent $2$ is smaller than the exponent ($\psi \approx 4 $) observed from direct simulations probably due to the mean-field approximation made in our theory and other higher order nonlinear effects.

\section{Preventing catastrophic forgetting by using landscape-dependent constraints} 

To demonstrate the utility of the theoretical insights gained so far, we tackle a long-lasting challenge in machine learning, i.e., how to prevent catastrophic forgetting (CF)~\cite{CF,Robins95}. After a deep neural network (DNN) learns to perform a particular task, it is trained for another task. Though the DNN can readjust its weights to perform well for the new task, it may forget the previous task and thus fail catastrophically for the previous task. 
To prevent forgetting, a recent study by Kirkpatrick et al~\cite{EWC} proposed the elastic weight constraint (EWC) algorithm to train a new task by enforcing constraints on individual weights based on their effects on the performance of the previous task. Here, armed with the new insights on the loss landscape and SGD dynamics, we propose to a landscape-dependent constraints (LDC) algorithm to train for the new task with constraints applied to the collective PCA modes from the previous task and with constraint strength determined by the loss landscape flatness of the previous task. 

Explicitly, when we find a solution $\vec{w}_1$ in the weight space for the first task (task-1) by SGD, we can also determine the loss function landscape around $\vec{w}_1$ in terms of the flatness parameter $F_{1i}$ along the $i-$th PCA direction $\vec{p}_{1i}$ for task-1, which can be obtained directly (cheaply) from the weight variance by using the variance-flatness relation Eq.~\ref{SF}. When learning the new task, we can use a modified loss function for the second task (task-2) by introducing an additional cost term that penalizes the network for going out of the attraction basin of the task-1 solutions: 
\begin{equation}
\tilde{L}_2(\vec{w}) =L_2(\vec{w}) + \lambda \sum_{i=1}^{N_c} \frac{((\vec{w}-\vec{w}_{1})\cdot \vec{p}_{1i})^2}{F_{1i}^2},
\label{L_2}
\end{equation}
where $L_2$ is the original loss function for task-2 and $N_c(\le N)$ is the number of constrained PCA modes from task-1. In general, constrains can be included for all previous tasks for sequential learning of more than two tasks.  
The overall strength of the constraints from task-1 is parameterized by $\lambda$. The strength of the constraint for a specific PCA mode depends inversely on the flatness of the loss function in that specific PCA direction. 

\begin{figure}[htbp]
\centering
\includegraphics[width=0.8\linewidth]{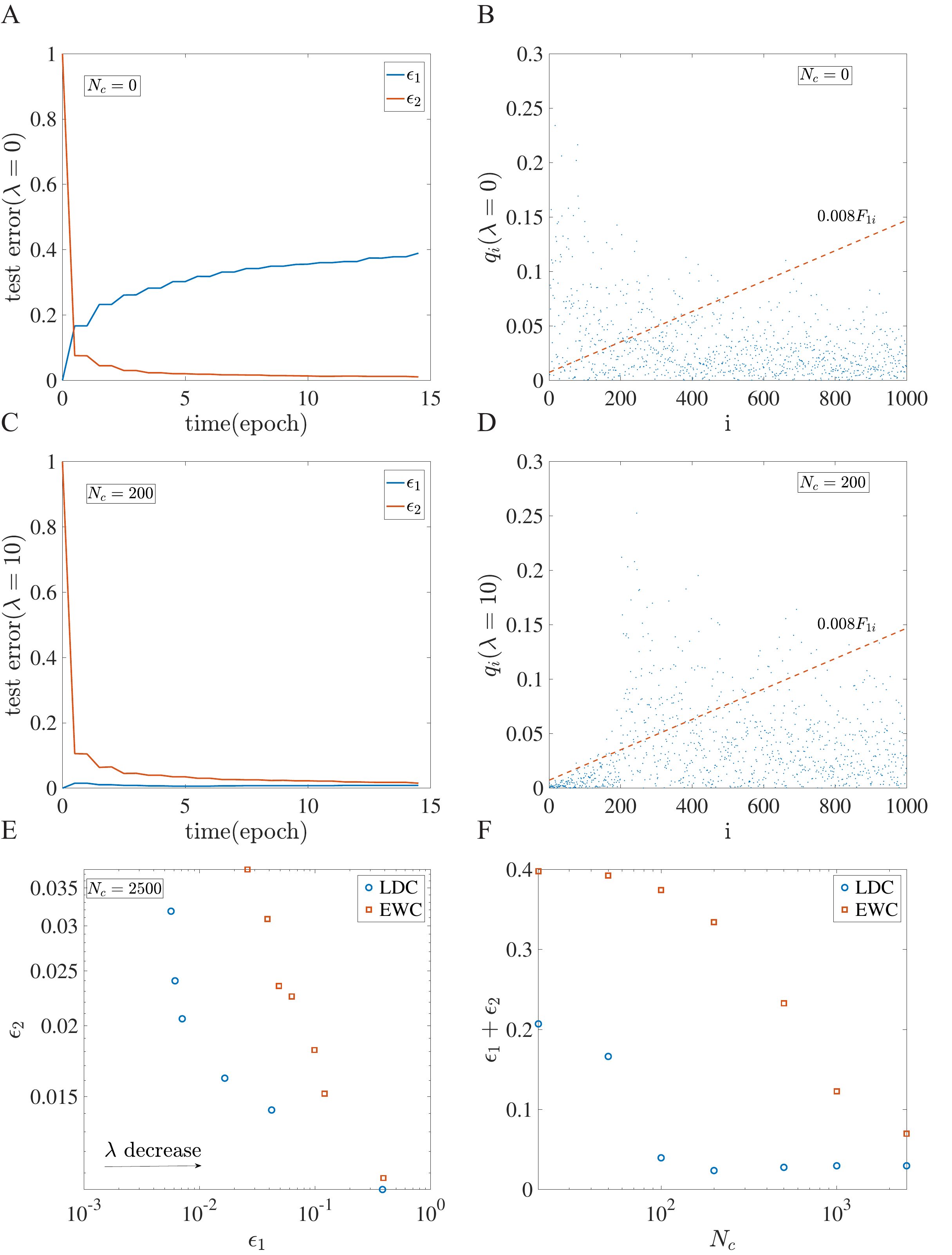}
\caption{The landscape-dependent constraints for avoiding catastrophic forgetting. (A) The test errors for task-1 ($\epsilon_1$) and task-2 ($\epsilon_2$) versus training time for task-2 in the absence of the constraints $(\lambda=0$). (B) The weight displacements $q_i$ in different PCA directions $\vec{p}_{1i}$ from task-1 in the absence of the constraints $(\lambda=0$). The threshold $\xi\equiv 0.008 F_{i1}$ is shown by the red dotted line. (C)\&(D) are the same as (A)\&(B) but in the presence of the constraints with $\lambda=10$ . (E) The tradeoff between the saturated test errors ($\epsilon_1$ and $\epsilon_2$) when varying $\lambda$ for LDC (blue circles) and EWC (red squares) algorithms. (F) The overall performance  ($\epsilon_1 +\epsilon_2$) versus the number of constrains $N_c$ for LDC (blue circles) and EWC (red squares) algorithms. The two tasks are for classifying two separate digit pairs [$(0,1)$ for task-1 and $(2,3)$ for task-2] in MNIST.}
\label{fig:4} 
\end{figure}

Based on the large attraction basin for a given task as evidenced by the large flatness parameters shown in Fig.~\ref{fig:2}, we expect that solutions for task-2 exist within the basin of solutions for task-1, so the performance for task-1 should not be degraded significantly after learning task-2. We tested this idea in the MNIST database with task-1 and task-2 corresponding to classifying two disjoint subsets of digits, e.g., (0,1) for task-1 and (2,3) for task-2 (see Methods for details).   
As shown in Fig.~\ref{fig:4}A in the absence of the constraints ($\lambda =0$), starting with a task-1 solution $\vec{w}_1$, the weights evolve quickly to a solution for task-2 $\vec{w}_2$ with a small task-2 test error $\epsilon_2$ (red line). However, the performance of task-1 deteriorates quickly with a fast increasing task-1 test error $\epsilon _1$ (blue line). The reason can be understood in Fig.~\ref{fig:4}B, where $q_i=||(\vec{w}_2-\vec{w}_1)\cdot \vec{p}_{1i}||$, the projections of the weight displacement vector onto different PCA directions of task-1, are shown. Without constraints, the displacement becomes much larger than a threshold $\xi_i$ set by the landscape flatness, $q_i\gg \xi_i(\equiv 0.008 F_{1i})$, for many high-ranking PCA modes (smaller $i$), which leads to the large task-1 test error after learning task-2, i.e., catastrophic forgetting.

The situation improves significantly when task-2 is learnt with the modified loss function $\tilde{L}_2$ given in Eq.~\ref{L_2}. As shown in Fig.~\ref{fig:4}C, in the presence of the constraints ($\lambda=10$) for the top $N_c=200$ modes, even though the learning process for task-2 is slightly slower, the system is able to learn a solution for task-2 with a comparable error $\epsilon_2$ as before ($\lambda=0$). The significant advantage here is that the performance for task-1 (blue line) remains roughly the same as before, i.e., the system has avoided catastrophic forgetting. The reason is explained in Fig.~\ref{fig:4}D, which shows that $q_i$ is bounded by the threshold ($\xi_i$), i.e., $q_i\lesssim \xi_i$, for the top modes ($i\le N_c$) due to the constraints. 

There is a tradeoff between the two testing errors ($\epsilon_1$, $\epsilon_2$) when varying $\lambda$. As shown in Fig.~\ref{fig:4}E, the performance of LDC is better than that of EWC. This is not surprising as LDC uses the full landscape information whereas EWC only uses the diagonal elements of the Fisher Information matrix (effectively the Hessian matrix). More interestingly, as shown in Fig.~\ref{fig:4}F, the overall performance ($\epsilon_1+\epsilon_2$) of LDC reaches its optimal level when a relatively small number ($N^*_c\approx 200$) of the top PCA modes are constrained. For EWC, however, all individual weights contribute to the performance, thus its optimal performance is reached when all $N=2,500$ individual weights are constrained. The results from the LDC algorithm suggest that memory of the previous task is encoded in the top $N_c^*$ PCA modes and $N^*_c$ can be used to estimate the capacity of the network for sequential learning.

\section{SGD as a self-tuned (landscape-dependent) annealing strategy for learning}

In the final section of our paper, we go back to evaluate our initial working hypothesis on the learning strategy deployed in SGD. In an equilibrium system with state variables $\vec{\theta}$ and a free energy function $L(\vec{\theta})$, the statistics of $\vec{\theta}$ follows the Boltzmann distribution $P(\vec{\theta})=\exp{[-L(\vec{\theta})/T]}$ where $T$ is the constant temperature that characterizes the strength of the thermal fluctuations (we set the Boltzmann constant $k_B=1$ here). By expanding the loss function to the second order: $L=L_{min}(1+\sum_i \frac{4 (\vec{\theta}\cdot\vec{p}_i)^2}{F_i^2})$ around a minimum $\vec{w}_0=0$, it is easy to show that the variance of $\theta_i$ would be proportional to the squared flatness $F_i^2$ and temperature $T$: \begin{equation}
\sigma^2_i \propto T\times F_i^2, 
\label{fdt}
\end{equation}
which is a direct consequence of the fluctuation-response (aka fluctuation-dissipation) relation in equilibrium statistical physics. 

Remarkably, for the SGD-based learning dynamics, we found an inverse relation between fluctuations of the variables and the flatness of the loss function landscape, Eq.~\ref{SF},
which is the opposite to the fluctuation-response relation in equilibrium systems given in Eq.~\ref{fdt}. We have tested it with different variants of the SGD algorithms such as ADAM and momentum based algorithms; different databases (MNIST and CIFAR-10); and different DNN architectures (see SM and Fig.~S3 for details). In all cases we studied, the inverse variance-flatness relation holds suggesting that it is an universal property of the SGD based learning algorithms. 

\begin{figure}[htbp]
\centering
\includegraphics[width=0.9\linewidth]{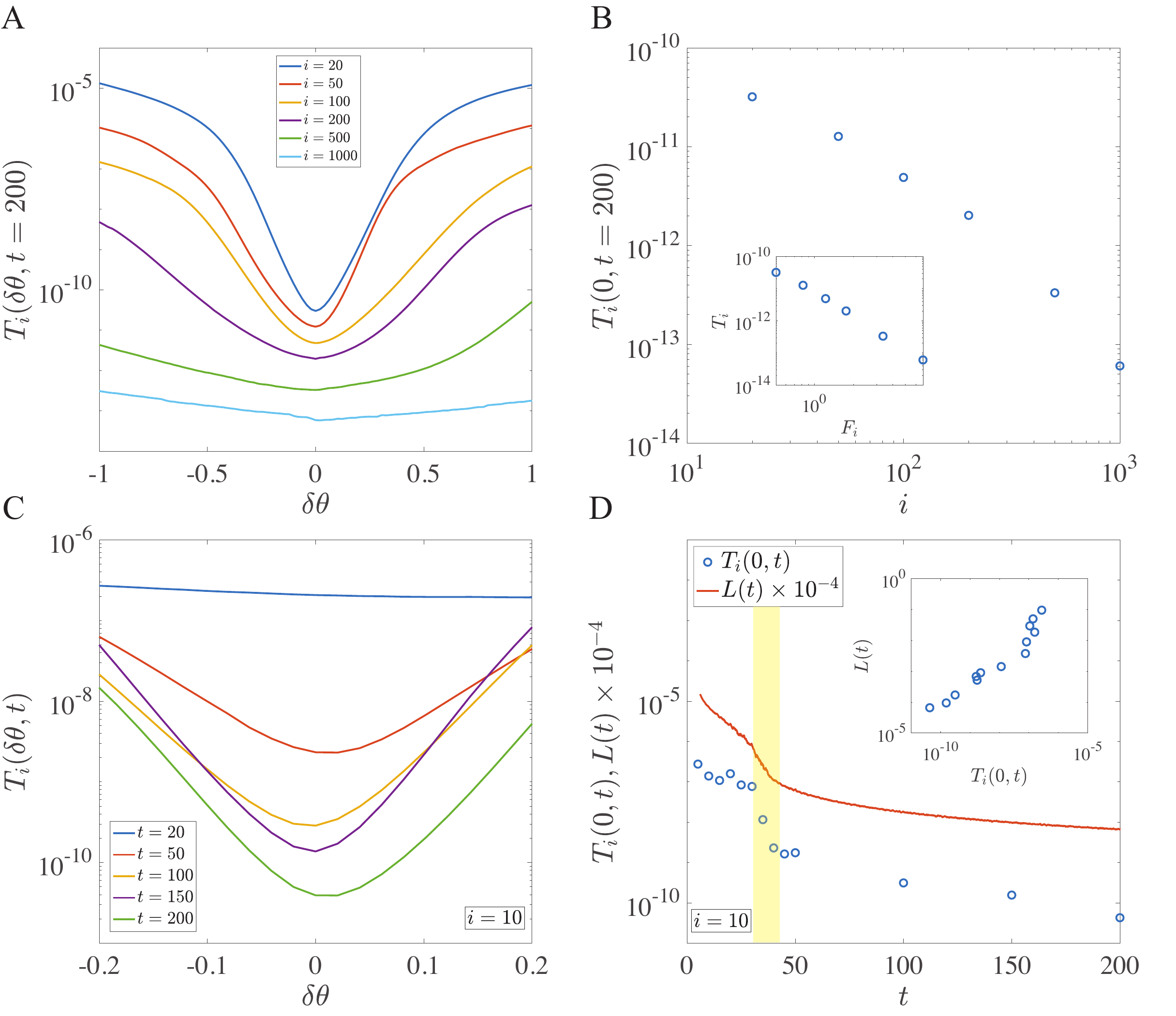}

\caption{Profiles and dynamics of the anisotropic active temperature. (A) The active temperature profile $T_i(\delta\theta,t)$ in the $i$'th PCA direction at $t=200$. (B) The minimum active temperature $T_i(0)$ in different PCA direction $i$. The inverse dependence of $T_i$ on the flatness $F_i$ is shown in the inset. (C) The active temperature profiles $T_i(\delta\theta,t)$ at different times for $i=10$. (D) The active temperature decreases with time in sync with the loss function (red line) dynamics. The shaded region highlights the transition between the fast learning phase and the exploration phase. The correlation between $T_i$ and $L$ is shown in the inset.} 
\label{fig:6} 
\end{figure}

Unlike thermal noise in equilibrium systems, which represents a passive random driving force with a constant strength (temperature), the SGD ``noise" $\vec{\eta}=\alpha\frac{\partial \delta L^\mu}{\partial \vec{\theta}}$ represents an active learning/searching process that varies in ``space" ($\vec{\theta}$). The intensity of this learning (search) activity -- learning intensity along the $i$-th PCA direction $\vec{p}_i$ can be characterized by an active local temperature $T_i(\delta\theta,t)$: 
\begin{equation}
T_i(\delta\theta,t)\equiv  \frac{\alpha}{2} \langle ||\frac{\partial \delta L^\mu (\vec{w}_0+\delta\theta \vec{p}_i)}{\partial \delta\theta}||^2\rangle_\mu, 
\end{equation}
where $\delta \theta$ is the displacement from $\vec{w}_0$ along the $\vec{p}_i$ direction. 

As shown in Fig.~\ref{fig:6}A, the active temperature $T_i(\delta \theta,t)$ has a similar spatial profile as that of the loss function with the active temperature higher away from the minimum. In weight space where the overall loss function is high, the active temperature is also high, which drives the system away from regions in the weight space with high losses.  The learning intensity also depends on the PCA direction as shown in Fig.~\ref{fig:6}B. For a flatter direction with a larger value of $F_i$, $T_i$ is lower (see the inset in Fig.~\ref{fig:6}B) as the basin of solutions is wide and thus no strong active learning is needed. However, for a steeper direction with a smaller value of $F_i$, the solutions exist only in smaller regions and thus more intensive learning (or higher active temperature) is required. Therefore, the MLF ensemble can sense both the local loss and non-local flatness of the landscape in different directions and use these information to drive active learning.

The active temperature also varies with time. As learning progresses, the active temperature profile decreases with time, as shown in Fig.~\ref{fig:6}C. In Fig.~\ref{fig:6}D, dynamics of the active temperature and the overall loss function along a SGD trajectory are shown together. It is clear that the active temperature and the overall loss function are highly correlated as shown directly in the inset of Fig.~\ref{fig:6}D, which means that the SGD system cools down as it learns. This reminds us of the well known simulated annealing algorithm for optimization~\cite{Kirkpatrick1983Optimization}, where temperature is decreased from a high value to zero with some prescribed cooling schedule. However, the SGD algorithm seems to deploy a more intelligent landscape-dependent annealing strategy where the active temperature (learning intensity), driven by the MLF ensemble, is self-tuned according to the local and non-local properties of the loss landscape that are sensed by the MLF ensemble. 
This landscape-dependent annealing strategy drives the system towards the flat minima of the loss function landscape and stays at the flat minima by lowering the active temperature once there. 

To summarize, a careful study of the SGD dynamics and the loss function landscape in this paper reveals a robust inverse relation between fluctuations in SGD and flatness of the loss landscape, which is critical for deciphering the learning strategy in deep learning and for designing more efficient algorithms. The statistical physics based approach provides a general theroretical framework for studying machine learning. We are currently using this approach to address other foundamental questions in machine learning such as generalization~\cite{Neyshabur2017ExploringGI,advani2017highdimensional}, relation between task complexity and network architecture, transfer learning~\cite{yosinski2014transferable}, and continuous learning~\cite{Ring94,GEM,riemer2018learning}.      

\section{Methods}

{\bf Neural network architecture and simulations.} Two types of DNNs are studied. (1) Two fully-connected Neural Networks were used for classifying digits in the MNIST database, one with two hidden layers ($784\times 50\times 50\times 10$, Main Text) and the other with four hidden layers ($784\times 50\times 50\times 50\times 50\times 10$, SM). The response of the hidden layer neurons is ReLu, activation of the output neurons is Softmax, and there is no bias neuron. We also studied Convolution Neural Network (CNN). (2) Two LeNets were used in our experiments. One is trained on MNIST dataset (see Fig.~S1 in SM) which has two convolution layer with size $1\times 3\times 3\times 16$ and $16\times 5\times 5\times 32$, and one fully-connected layer with size $1568\times 10$. (Here we represent the convolution layer using input neural number$\times$ kernel size$\times$ kernel size$\times$ output neural number).The stride of convolution is 1 and there is a zero padding to keep the data dimension unchanged. After each convolution layer, there is a $2\times 2$ max pooling layer.
Another CNN is trained on CIFA10 dataset (see Fig.~S3 in SM). It has has two convolution layer with size $3\times 5\times 5\times 6$ and $6\times 5\times 5\times 16$, and three fully-connected layer with size $400\times 120$, $120\times 84$, $84\times 10$. The stride of convolution is 1 and the size of max pooling is $2\times 2$. We do not use zero padding in this network. All numerical experiments are done on neural network simulation framework torch.

{\bf Principal component analysis in exploration phase.} For a given time in exploration phase, we extract the weight matrix ( $50\times50$) between two hidden layers and reshape the matrix to  $1\times2500$. Then we stack these row vectors from different times horizontally and did PCA on this matrix. The time step is each mini-batch and the total window size is $T=10$ epochs. The PCA is applied using sklearn package provided by Python 3.7.

{\bf Multi-task learning.} We divided the MNIST into 5 groups. Each group only contain 2 numbers. Here we call each group as task-1, task-2, etc. We use the fully-connected Neural Networks with two hidden layers ($784\times 50\times 50\times 10$). The size of output layer is 10 so it works for all tasks. For convenience, we choose the group containing (0,1) as task-1 and the group containing (2,3) as task-2. Firstly we train the network on all tasks. Then we initialize the hidden layer and fix all other layers. In this case, we only need to train the hidden layer. The multi-task learning process is described below: 1. train the network on task-1; 2. when the learning dynamics for task-1 reach the exploration phase, do PCA using method described in Part B; 3. train task-2 by using the modified loss function Eq.~\ref{L_2}. 

\section{acknowledgments}
We thank Mattia Rigotti, Irina Rish, Matthew Riemer, Robert Ajemain, Yunfei Teng, and Alberto Sassi for discussions. We also thank Jerry Tersoff, Tom Theis, and Youssef Mrouef for comments on the manuscript.  The work by Yu Feng was done when he was an IBM intern.

\bibliography{ML}

\end{document}


\title{Supplementary Material for ``How neural networks find generalizable solutions: Self-tuned annealing in deep learning"}
\maketitle

\section{The two phases of learning: the initial fast learning phase and the slow exploration phase}
The dynamics of $L$ versus iteration time are shown in Fig.~S1, which shows that there are two phases in learning in DNN. There is an initial fast learning phase where the overall loss function decreases quickly and sometimes abruptly followed by an exploration phase where the overall loss $L$ decreases much more slowly and gradually. Due to its slow time scale, the exploration phase can be considered as in quasi-steady-state. The weights reached in the exploration phase can be considered as solutions of the problem given their small loss function values. These two phases exist independent of the hyperparameters and the network architecture as shwon in Fig.~S1, where the transition region between the two phases are highlighted.

\section{Properties of the first principle component mode}

In our study, we used the cross entropy loss function for each sample:
\begin{equation}\tag{S1}
    d(\vec{Y}_k,\vec{Z}_k)=-\ln \big[\frac{\exp(\vec{Y}_k\cdot\vec{Z}_k)}{\sum_{n=1}^{n_c} \exp(Y_{k,n})}\big]=-\ln \big[\frac{\exp(Y_{k,n(k)})}{\sum_{n=1}^{n_c} \exp(Y_{k,n})}\big],
\end{equation}
where $n_c$ is the total number of classes and also the dimension of the network output vector $\vec{Y}_k$ and the correct output vector $\vec{Z}_k$. The correct class for sample $k$ is $n(k)$, and $\vec{Z}_k$ is just a unit vector in the correct ($n(k)$) direction.  Let $m(k)$ be the incorrect output component with the largest output value, we can then define $\Delta Y_k\equiv Y_{k,n(k)}-Y_{k,m(k)}$. The cross entropy loss for sample $k$ can then be written as:
\begin{equation} \tag{S2}
   l_k\equiv d(\vec{Y}_k,\vec{Z}_k)=\ln[1+a_k \exp(-\Delta Y_k)]\approx a_k \exp(-\Delta Y_k), 
   \label{dYZ}
\end{equation}
where $a_k=\sum_{n\ne n(k)} \exp(Y_{k,n}-Y_{k,m(k)})$ is an order O(1) number given that $Y_{k,m(k)}$ is the largest among all the incorrect  outputs ($n\ne n(k)$).

As we described in the main text, the persistent drift in the first PC direction $\vec{p}_1$ can be understood by translating a solution $\vec{w}_0$ found by SGD along $\vec{p}_1$ by $\theta_1$ to a new weight vector $\vec{w}=\vec{w}_0+\theta_1\times \vec{p}_1$. We studied how such a change of weights affects $\Delta Y_k$ for a given sample $k$. We computed $\Delta Y_k$ as a function of $\theta_1$ for different samples, and found that the change in $\Delta Y_k$ depends approximately linearly on $\theta_1$ :
\begin{equation} \tag{S3}
\Delta Y_k (\theta_1) \approx \Delta Y_k (0)+ \beta_k \theta_1,
\label{DY}
\end{equation}
where $\beta_k$ is a sample dependent constant. This linear dependence is shown clearly in Fig.~S2A for different samples covering different correct classes. 

By using Eq.~\ref{DY} in the expression for $l_k$, i.e., Eq.~\ref{dYZ}, we have:
\begin{equation} \tag{S4}
    l_k(\theta_1)\propto l_k(0)\times \exp(-\beta_k \theta_1), 
\end{equation}
which depends exponentially on $\theta_1$ for all samples $k$. As a result, the overall loss function along the first PC direction:
\begin{equation} \tag{S5}
    L_1\equiv \langle l_k (\theta_1)\rangle _k \approx L_0 \exp(-\beta \theta_1 +\beta_2 \theta_1^2),
    \label{L1}
\end{equation} 
where $\beta$ is the average of $\beta_k$: $\beta =\langle \beta_k \rangle _k>0$, and $\beta_2$ depends on the variation of $\beta_k$ among different samples. The expression for $L_1(\theta_1)$ agrees with the one obtained from the MLF ensemble average derived in the main text with the correspondences: $\beta=M_{11}^{(0)}\theta_1^{(0)}$ and $\beta_2=M_{11}/2$.

The functional form of $\ln(L_1(\theta_1))$ expressed in Eq.\ref{L1}, i.e., $ \ln(L_1(\theta_1)) \approx \ln(L_1(0))-\beta \theta_1 +\beta_2 \theta_1^2$ agrees with our numerical results shown in Fig.~S2B, which shows that $\ln(L_1)$ has a finite slope at $\theta_1$. It is this finite slope that drives the drift motion of $\theta_1$ observed in Fig.~1C in the main text.


\section{Robustness and generality of the inverse variance-flatness relation}

As shown in the main text (Fig.~2D), the inverse relation between the SGD variance and the flatness of the loss function landscape is the same for different choices of the hyperparameters, batch size $B$ and learning rate $\alpha$.
Since all the results reported in the main text are from a simple network with 2 hidden layers applied to the MNIST dataset, we have tested the robustness of this inverse variance-flatness relation in different networks, for different algorithms and datasets. We first tested the variance-flatness relation in a DNN with a larger number of hidden layers, e.g., 4 hidden layer network as shown in Fig.~S3A. We found the same inverse variance-flatness relation for weights in all different layers as shown in Fig.~S3B. We have also tested the inverse variance-flatness relationship in other DNN architectures such as LeNet (illustrated in Fig.~S3C), different learning algorithms such as ADAM, and different dataset such as CIFAR. As shown in Fig.~S3D, the inverse variance-flatness relation holds true in all these cases we studied.


\newpage

\begin{figure}[htbp]
\centering
\includegraphics[width=0.9\linewidth]{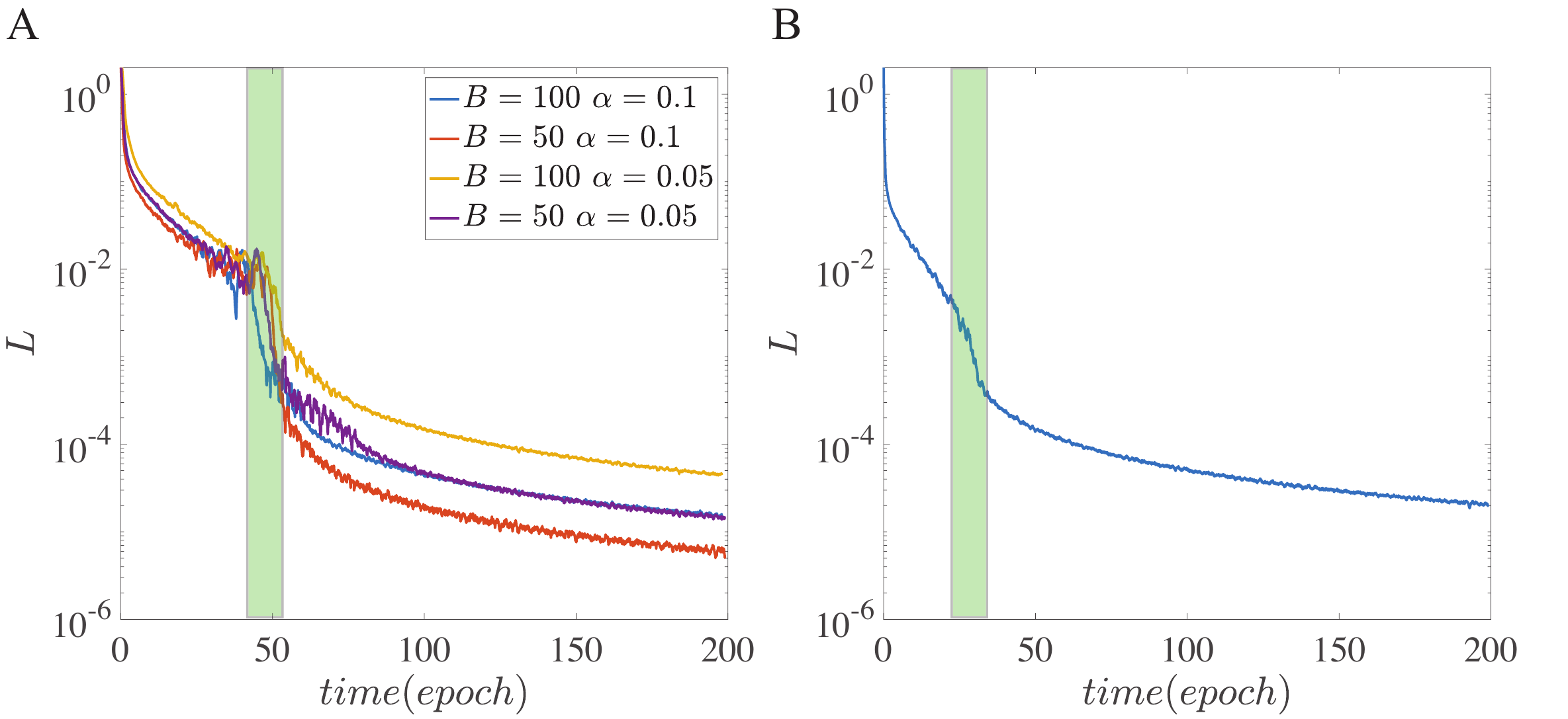}
\caption{Two different phases in learning. (A) shows the phase transition in a multi-layer fully-connected neural network. Here we use a network with four hidden layers and each hidden layer has 50 units. The experiment is done on the MNIST data sets. We can see that there is an obvious fast drop of cross-entropy loss around 50 epochs after that the system enters the exploration phase with slowly changing loss. This phase transition holds true for different combination of learning rate and batch size. (B) shows the phase transition in a LeNet, which has two convolution layers with size  $1\times3\times3\times16$ and $16\times5\times5\times32$, and one fully-connected neural network with size $1586\times10$ (see Fig.~S3C). We can see that there is also a drop of cross-entropy loss around 30 epochs}
\label{SM fig:1}
\end{figure}

\begin{figure}[htbp]
\centering
\includegraphics[width=0.9\linewidth]{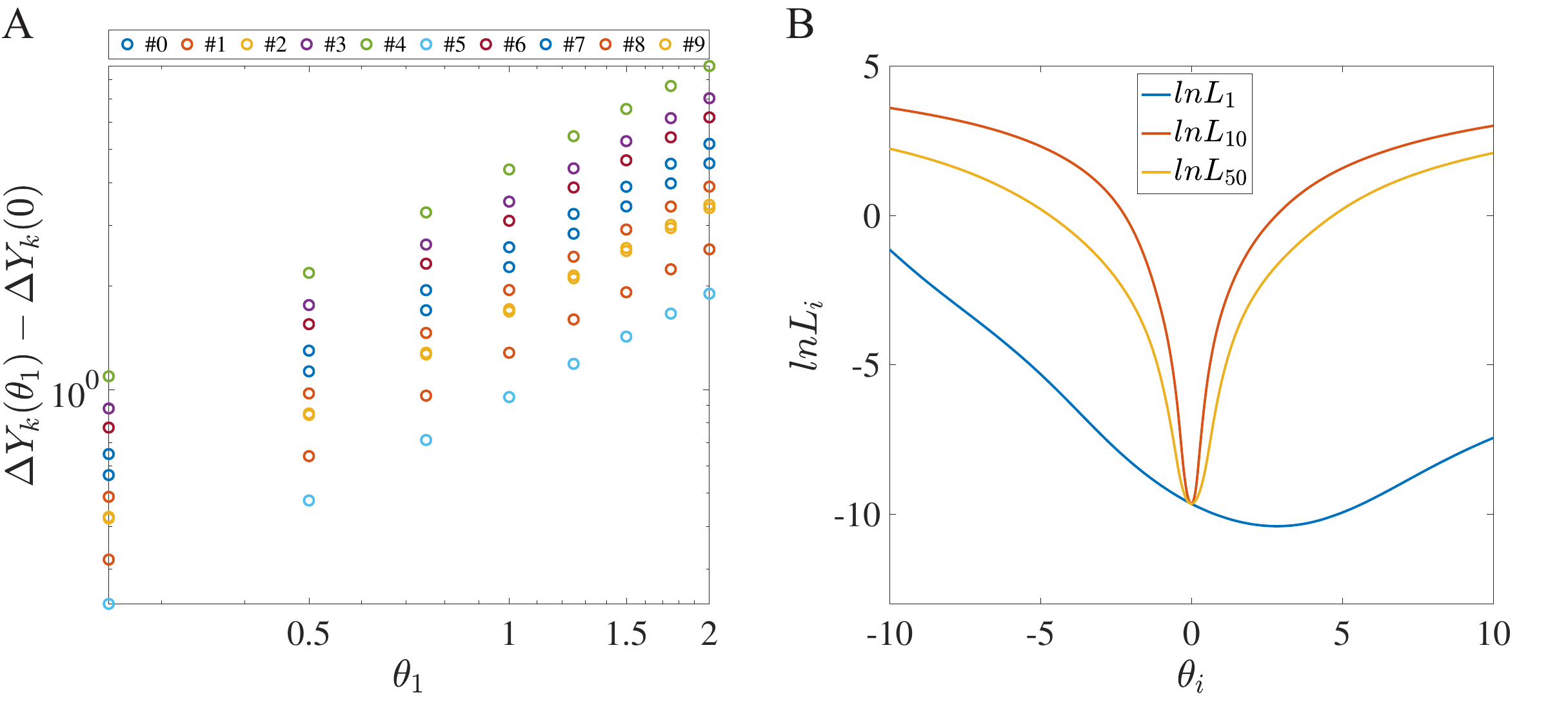}
\caption{The effect of changing $\theta_1$ along the first PC direction $\vec{p}_1$. (A) The dependence of $\Delta Y_k$, the difference between the correct output and the maximum incorrect output, on the displacement $\theta_1$ along the first PC for different sample $k$. $\Delta Y_k$ increases linearly with $\theta_1$. Each symbol corresponds to an average over $5$ samples of the same digit. (B) The landscape of $\ln L_i(\theta_i)$ along the $i$-th PC direction. Only $\ln(L_1)$ has a finite gradient at $\theta_1=0$, which indicates a drift motion in $\vec{p}_1$. In all other PC directions $(i\ge 2)$, $\ln(L_i)$ has a zero gradient at $\theta_i=0$, which indicates a diffusive motion in these directions.}
\label{SM fig:2}
\end{figure}

\begin{figure}[htbp]
\centering
\includegraphics[width=0.9\linewidth]{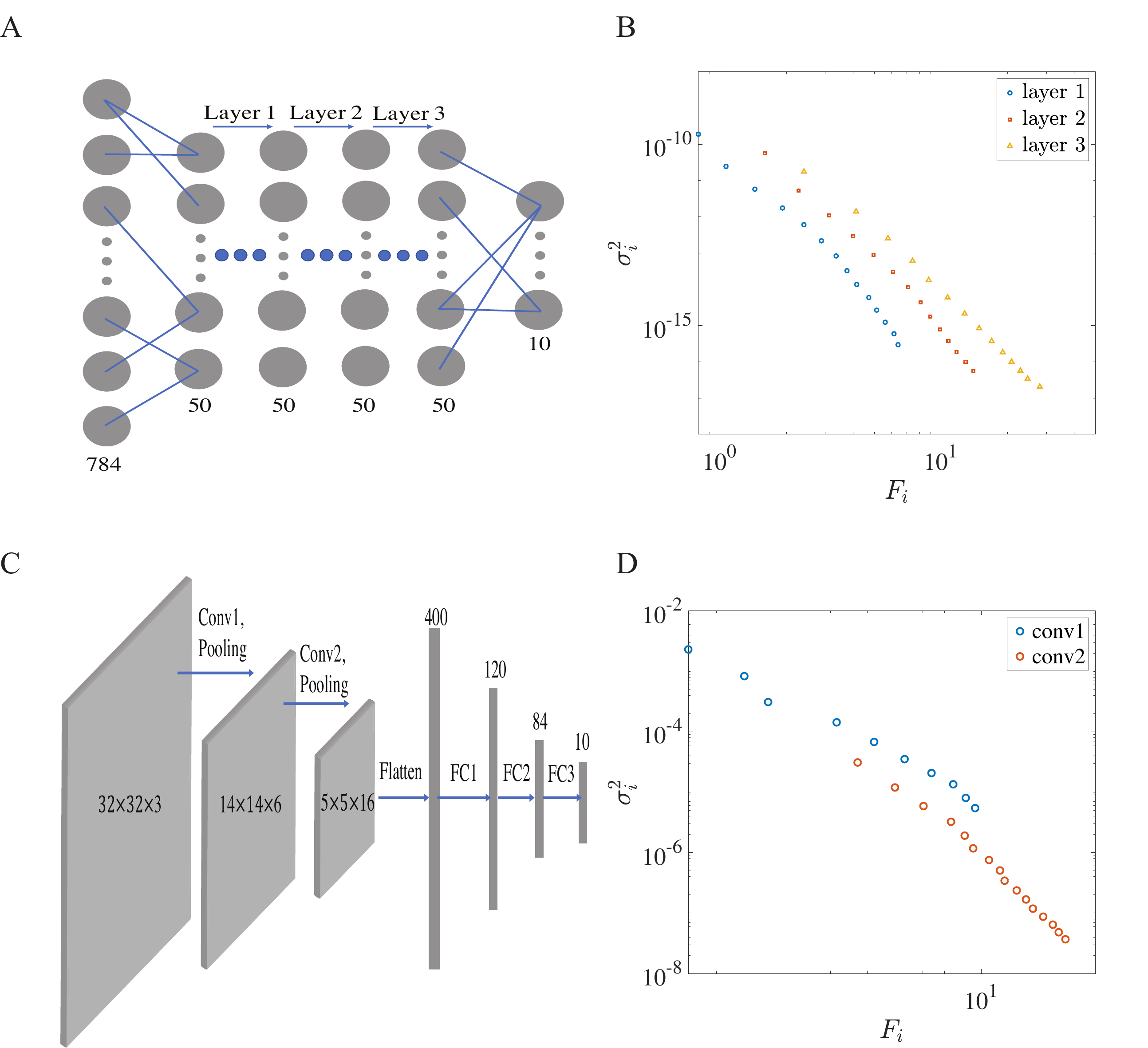}
\caption{The variance-flatness relation for different network, data set and optimization method. (A) A four hidden layer neural network is used on the MNIST dataset. (B) The inverse relation between variance and flatness holds between any two adjacent hidden layers. (C) Illustration of LeNet which has two convolution layer with sizes of $3\times5\times5\times6$ and $6\times5\times5\times16$, and three fully connected layers with sizes of $400\times120$,$120\times84$,$84\times10$. (D) The relation between variance and flatness in the two convolution layers clearly shows an inverse dependence. The LeNet convolution network shown in (C) is used on the CIFA10 dataset and the network is optimized by using SGD with momentum.}
\label{SM fig:3}
\end{figure}